\documentclass[epj,twocolumn]{webofc}
\usepackage[varg]{txfonts} 
\usepackage{epsfig}
\usepackage{wrapfig}
\usepackage{dcolumn}
\usepackage{bm}
\usepackage{subfigure}
\usepackage{longtable}
\usepackage{hyperref}
\usepackage{slashed}
\newcommand{\beq}{\begin{eqnarray}}
\newcommand{\eeq}{\end{eqnarray}}
\newcommand{\be}{\begin{equation}}
\newcommand{\ee}{\end{equation}}

\def\fun#1#2{\lower3.6pt\vbox{\baselineskip0pt\lineskip.9pt
\ialign{$\mathsurround=0pt#1\hfil ##\hfil$\crcr#2\crcr\sim\crcr}}}
\newcommand{\ppfrac}[2]{\dfrac{\partial{#1}}{\partial{#2}}}

\newcommand{{\SD}}{\rm SD}

\newcommand{{\Mc}}{\mathcal{M}}

\newcommand{\vep}{\mbox{\boldmath${\rm p}$}}
\newcommand{\veq}{\mbox{\boldmath${\rm q}$}}

\newcommand{\vecp}{\mbox{\boldmath${\rm p}$}}

\newcommand{\lbr}{\left(}
\newcommand{\rbr}{\right)}
\newcommand{\lbs}{\left[}
\newcommand{\rbs}{\right]}

\renewcommand\Im{\operatorname{Im}}
\begin{document}
%
%
\title{Sound Attenuation in Quark Matter Due to Pairing Fluctuations}

\subtitle{Talk at the 19th International Seminar ``Quarks -- 2016'', \\ Pushkin, Russia, 29 May -- 4 June, 2016}

\author{\firstname{Boris} \lastname{Kerbikov}\inst{1,2,3}\fnsep\thanks{\email{ borisk@itep.ru}}
}

\institute{{Alikhanov Institute for Theoretical and Experimental Physics} 
\and
           {Lebedev Physical Institute} 
\and
           {Moscow Institute of Physics and Technology}
          }

\abstract{
The sound  wave in dense quark matter is subject to strong absorption due to diquark field fluctuations above $T_c$. The result is another facet of Mandelshtam-Leontovich slow relaxation time theory.
}
\maketitle
%

During the last decade the investigation of quark matter at finite temperature and density became a compelling topic in QCD. Drawn in the $(T,\,\mu)$ plane with $\mu$ being the quark chemical potential the QCD phase diagram embodies several domains with quite different properties. Hopefully, we understand more or less the physics of quark-gluon matter in the high-temperature and low-density region. On the experimental side this is because most of the data on heavy-ion collisions obtained at RHIC and LHC correspond to high $T$ and zero, or very small, $\mu$. On the theoretical side the zero $\mu$ and high $T$ domain is accessible to Monte-Carlo simulations. 

Our focus in the present talk is on the opposite regime of high-density and moderate temperature. Such conditions may be realized in neutron stars and in future experiments at FAIR and NICA. The experimental data on this region are extremely scarce and theoretical understanding is on shaky ground. A powerful way to investigate the nature of a certain substance is to study its response to an external perturbation. This may be, e.g., magnetic field, temperature or pressure gradient, etc. Here we investigate the sound wave absorption in dense moderate temperature quark matter in the pre-critical region above $T_c$. According to the present understanding attractive interaction between quarks in the color antitriplet state leads at high density to the formation of the color superconducting phase. This phese is formed approximately in the region $\mu \gtrsim 400\,MeV$ and $T < T_c \sim 40\,MeV$. There are certain subtleties concerning the above numbers which are not crucial for our problem. There are two principal points we want to briefly elucidate in this short talk before the final results are presented. These two interrelated questions are:
\begin{itemize}
\item[i)] what is the difference between BCS and color superconductivity,
\item[ii)] what regime one has to attribute to the above domain of the QCD phase diagram: strong-coupling, weak-coupling, or BCS-BEC crossover.
\end{itemize} 
Both points were overwhelmingly discussed in literature including a few works of the present author \cite{01,02,03,04,05}. We consider the 2SC color superconducting phase when $u$- and $d$- quarks participate in pairing, while the density is not high enough to involve the $s$-quark in pairing. The very short answer to the question (i) is the following. Nonzero gap $\Delta$ is what BCS and 2SC have in common. In BCS two other important physical parameters are: the Debye frequency $\omega_D \sim 10^{-2}\,eV$ which measures the electron-phonon interaction and he Fermi energy $E_F \sim 2\,eV$. The BCS gap is $\Delta \sim T_c \sim \text{few}\,K$. The parameters of 2SC are: the ultra-violet momentum cuttoff $\Lambda \sim 0.8\,GeV$ and the chemical potential $\mu \sim 0.4 GeV$. The 2SC gap is $\Delta \sim 0.1\,GeV$. The hierarchy of the inherent scales in the two regimes is very different
\beq
\label{eq01}
BCS&\Delta:\omega_D:E_F \simeq 1:10^2:10^4,\\
\label{eq02}
2SC&\Delta:\Lambda:\mu\simeq 1:8:4.
\eeq
The fact that the BCS hierarchy is badly broken in 2SC reflects itself in the value of a dimensionless parameter $n^{1/3} \xi \sim k_F \xi$, where $n$ is the number density and $\xi$ is the characteristic length of pair correlation. In the BCS scenario $k_F \xi \gtrsim 10^3\,\text{\sc v.s.}\,k_F \xi \gtrsim 2$ in $2SC$ \cite{01,02,03,04,05}. In BCS pairs are large compared to their separarion, in 2SC they are small compared to their separation (such molecular-like objects are sometimes called Shafroth pairs). The continuous evolution of this parameter as a function of density and of the interaction strength reflects the transition from BCS regime to Bose-Einstein condensation (BEC). The idea of such transformation is due to F. Dyson, the problem was theoretically investigated in \cite{06} for electron pairs and following \cite{06} in \cite{02,03,04} for quarks. Starting from the early 2000th dramatic theoretical and experimental progress has been done in the study of the BCS-BEC crossover in ultracold fermionic atoms \cite{07}. 
  
Explaining the BCS-BEC crossover we partly elucidated the problem (ii) from the list above. A remark on terminology is needed. The terms strong- and weak-coupling in superconductivity (and in many-body physics) and in QCD have a different meaning. One can find, of course, some anology. Chiral symmetry breaking in QCD is a kind of a counterpart of the formation of fermion bilinear condensate in BCS. Roughly speaking, weak coupling in superconductivity means that the interaction between particles (electrons) is concentrated within a thin layer of momentum space around the Fermi surface and $\omega_D \ll E_F$ (see (\ref{eq01})). Integration in the vicinity of the Fermi surface is performed (in relativistic case) using the variable $\xi$ defined as 
\be\label{eq03}
\xi = \sqrt{\vecp^2 + m^2} - \mu,
\ee
\be\label{eq04}
\int\dfrac{d\vep}{(2\pi)^3}\simeq\int d\xi\,\rho(\xi)\simeq\int d\xi\lbs\rho(\mu)+\lbr\ppfrac{\rho}{\xi}\rbr_{\mu}\xi\rbs,
\ee
where $\rho(\mu)=\dfrac{p_0 \mu}{2 \pi^2}$, $p_0$ is the Fermi momentum. The first term in (\ref{eq04}) gives rise to Cooper logarithm. The second term takes into account the energy dependence of the density of states near the Fermi surface. Only this term in (\ref{eq04}) gives nonzero contribution to the sound absorption \cite{08,09}. In a great number of works the NJL (Nambu and Jona-Lasinio, Vaks and Larkin) model has been used to study color superconductivity. Strong-coupling in this approach is tantamaunt to a large value of the diquark coupling constant. In absence of a fundamental approach and impossibility to perform lattice calculations at nonzero density the NJL model may serve for the orientation purposes with a hope to confront its predictions with future FAIR and NICA data. As a candidate for a strong-coupling model the NJL suffers from a lack of renormalizability and confinement (the latter drawback is proposed to cure adding the Polyakov loop). 

On the other hand, the strong-coupling Migdal-Eliashberg theory of superconductivity was developed already more than half-century ago \cite{10,11} and has been successfully applied in many problems including HTSC \cite{12}.

Migdal-Eliashberg strong-coupling theory leads to the enhancement of pairing fluctuations and to the broadening of the transition region \cite{13}. Implementation of Migdal theorem \cite{11} to color superconductors will be discussed elsewhere. The fluctuation contribution is characterized by Ginzburg-Levanyuk number $Gi$ which for color superconductor becomes very large
\be\label{eq05}
Gi \simeq \dfrac{\delta T}{T_c} \simeq \lbr\dfrac{T_c}{\mu}\rbr^4 \simeq 10^{-4}.
\ee 
This is a huge number compared to the BCS $Gi \simeq 10^{-12}-10^{-14}$. An alternative estimate $Gi \sim (k_F \xi)^{-4} \sim 10^{-2}-10^{-3}$ \cite{04} leads to even larger value.

\begin{figure}[h]\centering
\begin{center}
\resizebox{0.70\columnwidth}{!}{\includegraphics[scale=1.0]{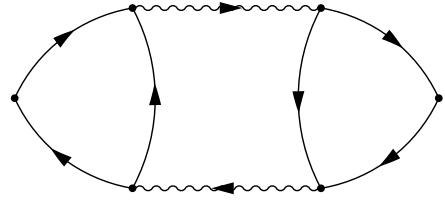}}
\end{center}
\vspace{-5 mm}
\caption{\label{fig:FIG1}Feynman diagram for the AL polarization operator for the sound absorption.}
\end{figure}

Precursor pair fluctuations above $T_c$ give the dominant contribution to the quark matter transport coefficients. The leading diagram is the Aslamazov-Larkin (AL) one \cite{08}, which includes two propagators of the slow pair collective mode singular at $T_c$. Within the ordinary theory of super conductivity this so-called fluctuation propagator (FP) was derived in \cite{08}. For color superconductor it was evaluated using Dyson equation and Matsubara formalism in \cite{14} and from the time-dependent Landau-Ginzburg equation with stohastic Langevin forces in \cite{09}. The FP reads   
\be\label{eq06}
L(\veq, \omega) = -\dfrac{1}{\nu}\,\dfrac{1}{\varepsilon +\dfrac{\pi}{8T_c}\lbr -i \omega + D\veq^2 \rbr}.
\ee
Here $\nu=\rho(\mu)=\dfrac{p_0 \mu}{2 \pi^2}$ (see (\ref{eq04})), $\varepsilon=\dfrac{T-T_c}{T_c}$, $D$ is the diffusion coefficient. At small $\omega$ and $q$ the quantity $L(\veq,\omega)$ can be arbitrary large close to $T_c$. The AL diagram for the sound absorption is shown in Fig.\ref{fig:FIG1}
\vspace{-5 mm}
The two way lines correspond to the FP-s, the solid lines are the quarks Matsubara propagators, the in- and out- vertices are equal to the constants $g$ of phonon-quark interaction. The sound attenuation coefficient is equal to the imaginary part of the retarded polarization operator corresponding to the diagram in Fig.\ref{fig:FIG1}. In this short presentation we leave out the detailed calculation which may be found in \cite{08} for BCS and in \cite{09} for the quark matter. In the fluctuation region above $T_c$ the dominant contribution and the sharp temperature dependence come from the two FP-s. The final result for the imaginary part of the polarization operator reads \cite{09}
\be\label{eq07}
\begin{split} \Im \Pi =& -\omega\,g^2\,\dfrac{m^2}{2^5\,p_0^4\,\varkappa^3}\,(v_0^2 + 1)^2\times \\ \times &\ln^2 \frac{\Lambda}{2 \pi\,T_c} \lbr\dfrac{T_c}{T-T_c}\rbr^{3/2}. 
\end{split}
\ee
Here $m$ is the quark mass, $p_0$ and $v_0$ are the Fermi momentum and velocity $\Lambda$ is the ultra-violet cutoff (\ref{eq02}), $\varkappa^2=\dfrac{\pi}{8T_c}D$, where $D$ is the diffusion coefficient. As we remarked after Eq.(\ref{eq04}), the nonzero contribution to $\Im \Pi$ comes not from crust of the Fermi surface (the first term in (\ref{eq04})), but from the energy dependent density of states given by the second term in (\ref{eq04}). Important to note that the evaluation of the AL diagram in \cite{08} and \cite{09} making use of (\ref{eq04}) are in complete agreement with the result for this polarization operator obtained in the Eliashberg strong-coupling theory \cite{13}. 

The physics behind the strong energy dissipation of the sound wave in the precritical region is simple and very general. This is another manifestation of Mandelshtam-Leontovich slow relaxation time theory \cite{15,16,17}. Propagation of the sound wave locally changes the critical temperature and the slow fluctuation pairing can not keep up with this process. 

The author is supported by a grant from the Russian Science Foundation number 16-12-10414. The author express his gratitude for discussions and remarks to A.~Varlamov, Yu.~A.~Simonov, M.~Lukashov, M.~Andreichikov, L.~McLerran, P.~Petreczky.

%
%
%
%

\end{document}